# Investigation of the gamma-ray shielding performance of the $B_2O_3$-$Bi_2O_3$-ZnO-$Li_2O$ glasses based on the Monte Carlo approach

Ali Asadi[a], Seyed Abolfazl Hosseini[a,1]

[a] *Department of Energy Engineering, Sharif University of Technology, Tehran, Iran, Zip code: 8639-11365*

**Abstract:**

The purpose of this article is to investigate the shielding performance of the $B_2O_3$-$Bi_2O_3$-ZnO-$Li_2O$ glasses as gamma shields. To this end, the attenuation parameters of the gamma-ray for $B_2O_3$-$Bi_2O_3$-ZnO-$Li_2O$ glasses were calculated from the results of the simulation performed by MCNPX computer code. To validate the simulation, the calculated values of mass attenuation coefficients in the energy range of 200 keV to 1500 keV were compared with the XCOM data base. The relative deviation between the results of simulation using the MCNPX and the XCOM database was 2%. Additionally, the mean free path (MFP) and half-value layer (HVL) parameters were calculated. The results show that among the examined samples, the $B_4$ glass sample has the best shielding performance.

From the results of the calculation, it can be understood that the addition of compound $Li_2O$ to compound $B_2O_3$-$Bi_2O_3$-ZnO-$Li_2O$ has a huge impact on the shielding performance of the examined glass versus gamma-rays. In addition, the results show that the existing $B_2O_3$-$Bi_2O_3$-ZnO-$Li_2O$ glasses will have a promising outlook as gamma rays shield due to the possibility of changing the weight percentage of $Li_2O$ in them.

1Corresponding author. Tel.:+98 21 6616 6140; fax:+98 21 66081723
E-mail address: sahosseini@sharif.edu (Seyed Abolfazl Hosseini).

# 1. Introduction

In recent years, high energetic gamma rays have been used in various applications such as response of some electronic devices to radiation, material analysis, radiation of human consumables, advanced scientific research, etc.(Kahraman and Yilmaz, 2017; Kahraman et al., 2016). Because of the high penetrating power of these rays, these have a big worry for human health and environmental safety, for example, electronic laboratory equipment (Sayyed et al., 2018a). As is known, designing a suitable radiation shield has the greatest effect on reducing radiation damage. Therefore, determining the parameters related to the beam passage through the absorber has always been considered. Because the addition of lead to any substance causes toxicity in the nature of that substance (Agar, 2018), excessive use of lead is not appropriate. In contrast, concrete materials that are very cheap and available are widely used to protect gamma rays (Kumar et al., 2019). However, ordinary materials have fundamental problems:(1) Concrete is opaque to visible light, (2) They are difficult to move due to their heavyweight (Ashok et al., 2018). Today, glass materials are receiving more attention as new and better protective materials. Recently, much research has been done on the radiation protection functions and the structural, mechanical, and physical properties of various glass systems (Issa et al., 2018; Sayyed et al., 2018b). Despite the efficiency of experimental methods to validate scientific work, today, the use of computer-based computational and numerical methods to solve physical problems is increasing continual. For example, the Monte Carlo simulation method provides a suitable laboratory environment taking into account all physical conditions such as the geometry of the structure of radiation sources, etc. Further, the advantage of application of such a tool during calculations is related to the radiation protection of the material, reducing the study time, minimizing the study cost, and reducing the exposure time of the experimenter. That's why we chose one of the famous Monte Carlo codes, MCNPX 2.7, to study the protective properties of a particular type of glass.

Many researchers have tried to simulate the gamma-ray shielding parameters of composite shields. Some of these researches have been selected and briefly discussed below:

1- Sharma and his colleagues used FLUKA computer code to simulate the radiation shielding parameters of $TeO_2$-$WO_3$-$GeO_2$ glasses, such as MFP, HVL, and mass attenuation coefficient. The results from FLUKA were compared to the XCOM database values. They found that FLUKA computer code has an excellent capability to simulate the radiation shielding parameters. According to the results, the radiation shielding parameters of composition with high $Z$ (atomic number), and high density elements have a superior protection performance (Sharma et al., 2019).

2 -Bagheri and his colleagues simulated the radiation shielding parameters of barium-bismuth-borosilicate glasses. The transmission factor, HVL, TVL, and mass attenuation coefficient were simulated in this work. They found that the glasses with high $Z$ (atomic number), and high density elements have better protective properties (Bagheri et al., 2017).

3- Akman and his colleagues experimentally investigated the protective performance against gamma-ray for Fe/Cr/Ni s like. The parameters like HVL, TVL, MFP, linear and mass attenuation, effective conductivity, and effective atomic number were calculated and validated against the data obtained from winXCOM database. They observed that experimental data are close to the data obtained from winXCOM's data. They found that the radiation shielding performance of Fe/Cr16/Ni72 is excellent (Akman et al., 2019).

In continuation of previous studies for simulation and validation of gamma-ray shielding performance of any composition samples, the purpose of this study is the investigation of the radiation shielding performance of a new shield material composed of B, Bi, Zn, Li and O element against gamma-ray. To this end, the gamma-ray shielding parameters such as MFP, HVL, and mass and linear attenuation coefficient of the $B_2O_3$-$Bi_2O_3$-$ZnO$-$Li_2O$ shield are calculated via Monte Carlo simulation. The results of simulation are validated against the data obtained from the XCOM database.

The remaining outlines of this study are as follows: In Section 2, we present the materials used in this study and methodology of theoretical calculation of gamma-ray shielding factors of $B_2O_3$-$Bi_2O_3$-$ZnO$-$Li_2O$ samples. In Section 3, numerical results and discussion on results are presented based on the results obtained from the MCNPX simulation and data obtained from XCOM database for the gamma-ray shielding parameters of the mentioned shield. Finally, In the Section 4, a summary of the study and conclusion is reported.

## 2. Materials & methods

### 2.1. Geometry of glass sample

In this work, cubic geometry was used to model $B_2O_3$-$Bi_2O_3$-$ZnO$-$Li_2O$ glass. The number of 5 sections inside each cube with dimension of ~~1 cm×1 cm~~ was considered.

### 2.2. Source specification

To calculate the mass attenuation coefficient of the glass sample, a mono-energetic point source of the gamma-ray was used. The point source was positioned at a distance of 10 cm from the glass sample, where the gamma-ray strikes the front surface of the shield vertically (Fig. 1). In MCNPX computer code, this source was defined by data card, with POS, PAR, ERG, VEC, and DIR command for position, particle kind, energy, direction vector, and direction, respectively.

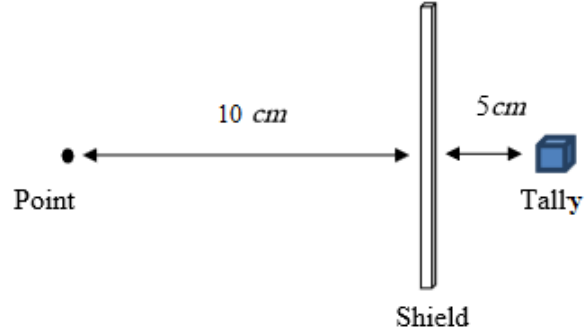

**Fig. 1.** Geometry of modeled configuration.

### 2.3. Material characteristics of samples

The main composition of each sample depends mainly on the chemical composition and the proportion of used materials. Based on the early researches, the composition of the glass sample was considered as $15B_2O_3$-$50Bi_2O_3$-$(35-x)$ ZnO-xLi$_2$O borosilicate glasses; where, x is expressed as mole percentage (x is equal to 15, 10, 5, 0). The chemical composition and density of each examined sample are reported in Table 1. In addition, percentage weight of each used material in the sample is presented in Table 2.

**Table 1.** The density and composition of the $B_2O_3$-$Bi_2O_3$-ZnO-Li$_2$O glass samples.

| Glass Sample | Sample composition | Density (g/cm³) |
|---|---|---|
| B₁ | 15B$_2$O$_3$-50Bi$_2$O$_3$-35ZnO | 5.98 |
| B₂ | 15B$_2$O$_3$-50Bi$_2$O$_3$-30ZnO-5Li$_2$O | 5.83 |
| B₃ | 15B$_2$O$_3$-50Bi$_2$O$_3$-25ZnO-10Li$_2$O | 5.66 |
| B₄ | 15B$_2$O$_3$-50Bi$_2$O$_3$-20ZnO-15Li$_2$O | 5.45 |

**Table 2.** The weight percentage of each element in the investigated B$_2$O$_3$-Bi$_2$O$_3$-ZnO-Li$_2$O glass system

| Elements | B1 | B2 | B3 | B4 |
|---|---|---|---|---|
| Li | 0.0 | 0.023228 | 0.046457 | 0.069685 |
| O | 0.223722 | 0.240664 | 0.257607 | 0.274549 |
| B | 0.046586 | 0.046586 | 0.046586 | 0.046586 |
| Zn | 0.281198 | 0.241027 | 0.200855 | 0.160684 |
| Bi | 0.448495 | 0.448495 | 0.448495 | 0.448495 |

## 2.4. Theoretical basis

In this study, different shielding parameters such as the HVL, MFP, and the mass attenuation coefficient of B$_2$O$_3$-Bi$_2$O$_3$-ZnO-Li$_2$O glass systems was determined using the MCNPX computer code. The results were reported for gamma-rays with energies of 200, 400, 800, 1000 and 1500 keV. To determine the efficiency of Li$_2$O on the gamma-ray shielding performance of B$_2$O$_3$-Bi$_2$O$_3$ -ZnO-Li$_2$O glass samples, the compositions, namely 15B$_2$O$_3$-50Bi$_2$O$_3$-35ZnO, 15B$_2$O$_3$-50Bi$_2$O$_3$-30ZnO-5Li$_2$O, 15B$_2$O$_3$-50Bi$_2$O$_3$-25ZnO-10Li$_2$O, and 15B$_2$O$_3$-50Bi$_2$O$_3$-20ZnO-15Li$_2$O were considered (Table 1). In the middle energy range, the intensity of the beam decreases as it passes through the material due to various interactions that occur when the beam passes through the material, such as photoelectric, Compton, etc. The intensity reduction is given by Lambert-Beer's law as Eq. (1) (Agar et al., 2019a; Akman et al., 2019):

$$I = I_0 \exp(-\mu x) \quad (1)$$

Where 'µ' is the linear attenuation coefficient of the absorber. The $I_0$ and $I$ represent the input intensities and the intensities transmitted from the shield, respectively. Also, 'x' is the diameter (cm) of the glass sample. Because the linear attenuation coefficients do not depend on the physical state of the material (solid, liquid, gas) and the density of the absorbent material, the mass attenuation coefficient parameter is much more important than the linear attenuation coefficient. The mass attenuation coefficient ($\mu_m$) is obtained from the Eq. (2) (Eke et al., 2017):

$$\mu_m = -\frac{1}{\rho x} \ln\left(\frac{I}{I_0}\right) \quad (2)$$

where, the parameter $\rho$ represent density (g/cm³) of the sample. By sketch $ln(I_o/I)$ vs. thickness of the sample, the slope of the curve determines the value of the mass attenuation coefficient "$\mu_m$".

Another important parameter in the radiation shielding is MFP. It is defined as the average distance between two consecutive interactions of a photon beam in matter in which the intensity of the collision photon beam decreases by 1/e factor. The MFP (cm) can be calculated by considering the linear attenuation coefficient ($\mu$) as Eq. (3)(Akman et al., 2019):

$$\lambda_{mean} = = \frac{1}{\mu} \quad (3)$$

In addition, another parameter that is of interest discussions related to radiation shielding is the HVL. It is defined as the thickness required to decrease the amount of radiation intensity to half the initial intensity amount. HVL may be calculated from Eq. (4) (Agar et al., 2019b; Alajerami et al., 2020):

$$HVL = \frac{\ln 2}{\mu} \qquad (4)$$

Knowing the worth of MFP, and HVL help to quickly evaluate radiation shielding performance of any material.

## 3. Results and discussion

### 3.1. Validation

In this study, the values of $\mu_m$ at five energies were calculated using the MCNPX computer code. Also, the same parameter was obtained from the XCOM program in the considered energies. The geometry expressed in section 2.2 was simulated using MCNPX computer code, and the $\mu_m$ parameter was evaluated for $B_2O_3$-$Bi_2O_3$-$ZnO$-$Li_2O$ glasses at the same energies. The results are reported in table 3 and graphically shown in Figs. 2(a-d). Moreover, the Relative percentage difference between MCNP and XCOM database were calculated according to Eq. (5):

$$\text{Relative Percentage Error} = \frac{(\mu_{m,MCNP} - \mu_{m,XCOM})}{\mu_{m,XCOM}} * 100 \qquad (5)$$

Table 4 indicates the comparison between the simulated $\mu_m$ values obtained from the simulation using MCNPX computer code and XCOM database. As shown in Table 4 and Figs. 2-5, the simulated $\mu_m$ values by MCNPX computer code are in good agreement with the values obtained from the XCOM database. Also, the range of Relative Percentage Error (RPE) obtained between the two methods is [1–6%] (for $B_1$), [0.37-6%] (for $B_2$), [0.33–6%] (for $B_3$) and [0.05–7%] (for $B_4$). The calculated small values of RPE for all cases in the energy range examined in this work indicate that the values of $\mu_m$ can be estimated for the $B_2O_3$-$Bi_2O_3$-$ZnO$-$Li_2O$ samples as well as for other samples. For the $15B_2O_3$-$50Bi_2O_3$-$35ZnO$ sample, the data obtained from the simulation (MCNPX) and measured (XCOM) results of $\mu_m$, are plotted in Fig. 2-a. Correlation theory was used to examine the linearity of the relationship between the two data sets )MCNPX and XCOM database results(. The correlation coefficient (r=0.9998) indicates that there is a high degree of agreement between the data obtained from the simulation and XCOM data.

| Glass Sample | 200keV | | 400keV | | 800keV | | 1000keV | | 1500keV | |
|---|---|---|---|---|---|---|---|---|---|---|
| | MCNP | XCOM | MCNP | XCOM | MCNP | XCOM | MCNP | XCOM | MCNP | XCOM |
| $B_1$ | 0.5748 | 0.5419 | 0.1632 | 0.1596 | 0.0775 | 0.0780 | 0.06529 | 0.0606 | 0.05173 | 0.05113 |
| $B_2$ | 0.5751 | 0.5399 | 0.1638 | 0.1593 | 0.0774 | 0.07810 | 0.0653 | 0.06461 | 0.0513 | 0.05111 |
| $B_3$ | 0.5721 | 0.5380 | 0.1637 | 0.1590 | 0.07767 | 0.07805 | 0.06578 | 0.06600 | 0.05171 | 0.05108 |
| $B_4$ | 0.5752 | 0.5361 | 0.1640 | 0.1587 | 0.07804 | 0.07800 | 0.06492 | 0.06597 | 0.05220 | 0.05106 |

**Table 3.** Mass attenuation coefficients (cm$^2$/g) of $B_2O_3$-$Bi_2O_3$-ZnO-$Li_2O$ glass samples

**Table 4.** RPE between MCNP and XCOM results.

| Glass Sample | 200keV | 400keV | 800keV | 1000keV | 1500keV |
|---|---|---|---|---|---|
| $B_1$ | 0.06 | 0.02 | 0.03 | 0.01 | 0.01 |
| $B_2$ | 0.06 | 0.028 | 0.03 | 0.01 | 0.0037 |
| $B_3$ | 0.06 | 0.029 | 0.005 | 0.0033 | 0.01 |
| $B_4$ | 0.07 | 0.033 | 0.0005 | 0.015 | 0.02 |

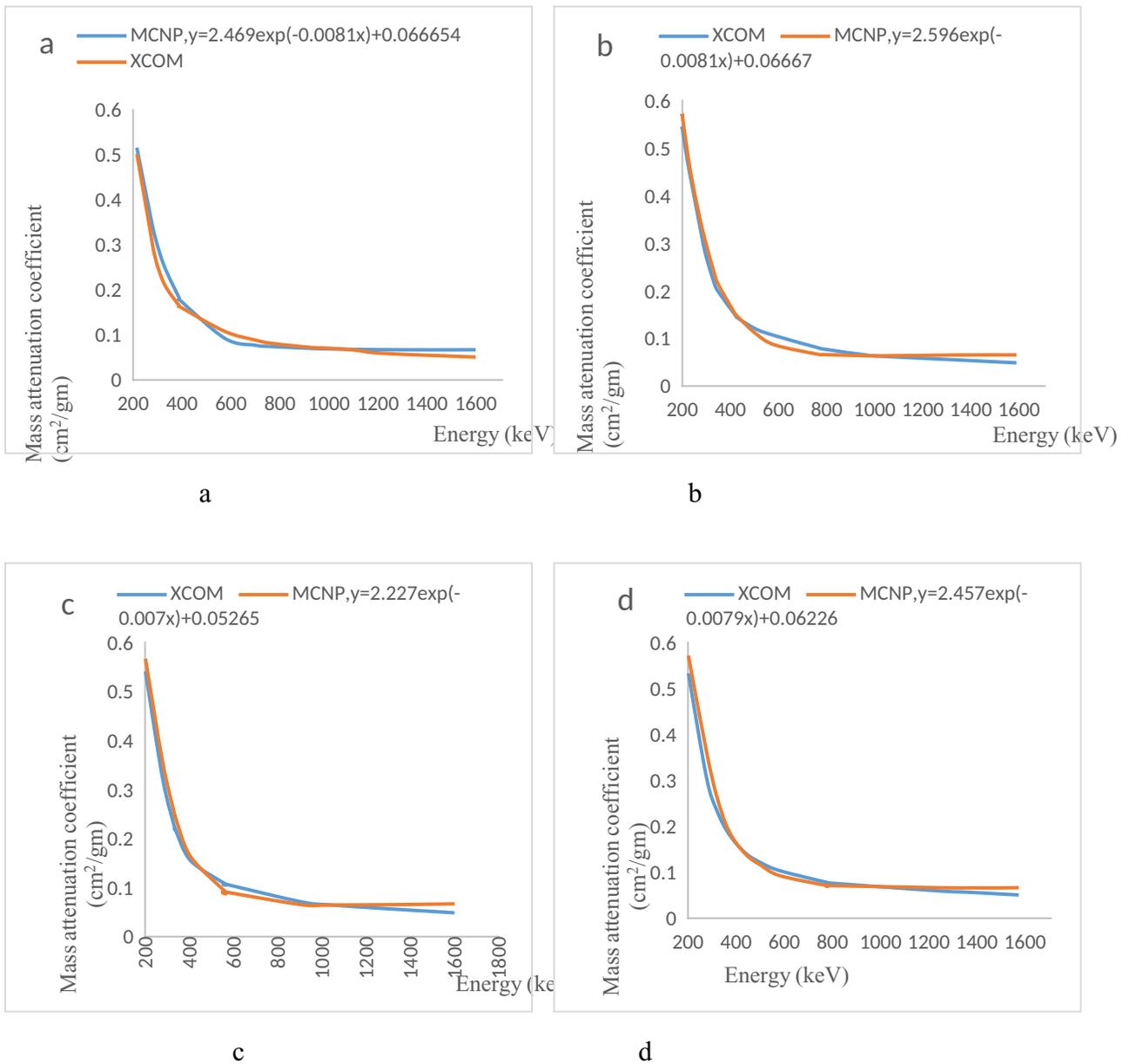

**Fig. 2.** The changes of mass attenuation coefficient of glass system vs. gamma energy for (a) $B_1$, (b) $B_2$, (c) $B_3$, and (d) $B_4$.

## 3.2 Attenuation properties of $B_2O_3$-$Bi_2O_3$-ZnO-$Li_2O$ glass system

In the following, the MCNPX computer code is used to calculate the $\mu_m$ values for $B_2O_3$-$Bi_2O_3$-ZnO-$Li_2O$ system to determine the efficacy of $Li_2O$ on the gamma-ray shielding performance of this samples. The composition, density, and weight fraction of each element of the samples are shown in Tables 1 and 2. The passage fraction (logarithmic scale) vs. the diameter (linear scale) of samples at various energies are graphically shown in Figs. 3(a-d) and Tables 5(a-d). From the Lambert-Beer's law (Eq. (1)), it is easy to understand that the slope of the line in fitted equations (obtained from the plot of $\ln(I/Io)$ vs. thickness) indicate linear attenuation coefficient. The presence of a negative sign in these equations indicates that the passing fraction decreases with increasing thickness of glass samples. In order to have MFP that is independent from the physical nature (density) of the material, the linear attenuation coefficient is obtained from these equations of sample material. From Table 5, for $15B_2O_3$-$50Bi_2O_3$-35ZnO in the energy of 662 keV, the linear fitting equation is y=−0.542x−0.0386; where the slope of this line is 0.542. This slope value indicates the linear attenuation coefficient ($cm^{-1}$) of the sample. By dividing the value of this parameter to the density of the glass sample (5.988 g/$cm^3$), the value of MFP is calculated as 0.0906 $cm^2$/g.

**Table 5.** Fitted equations for variation of intensity of gamma photons vs. thickness of sample for (a) $B_1$, (b) $B_2$, (c) $B_3$, and (d) $B_4$.

a

| Energy(keV) | Fitted equation |
|---|---|
| 357 | y = − 1.157x + 0.0142 |
| 662 | y = − 0.542x + 0.0366 |
| 1173 | y = − 0.3608x + 0.07525 |
| 1330 | y = − 0.3258x + 0.04373 |

b

| Energy(keV) | Fitted equation |
|---|---|
| 357 | y = − 1.136x + 0.02756 |
| 662 | y = − 0.5436x + 0.06407 |
| 1173 | y = − 0.354x + 0.06237 |
| 1330 | y = − 0.3288x + 0.05864 |

c

| Energy(keV) | Fitted equation |
|---|---|
| 357 | y = − 1.098x + 0.02034 |
| 662 | y = − 0.5303x + 0.0661 |
| 1173 | y = − 0.3425x + 0.05124 |
| 1330 | y = − 0.3203x + 0.0561 |

d

| Energy(keV) | Fitted equation |
|---|---|
| 357 | y = − 1.055x + 0.01763 |
| 662 | y = − 0.5124x + 0.06871 |
| 1173 | y = − 0.329x + 0.06206 |
| 1330 | y = − 0.2758x − 0.004746 |

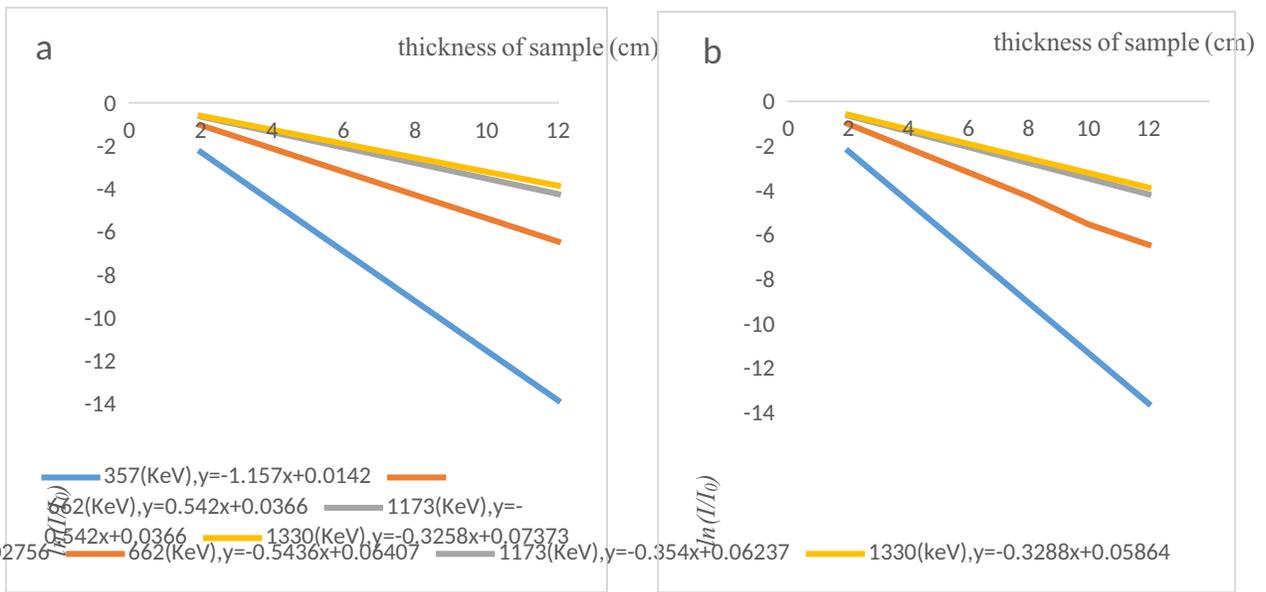

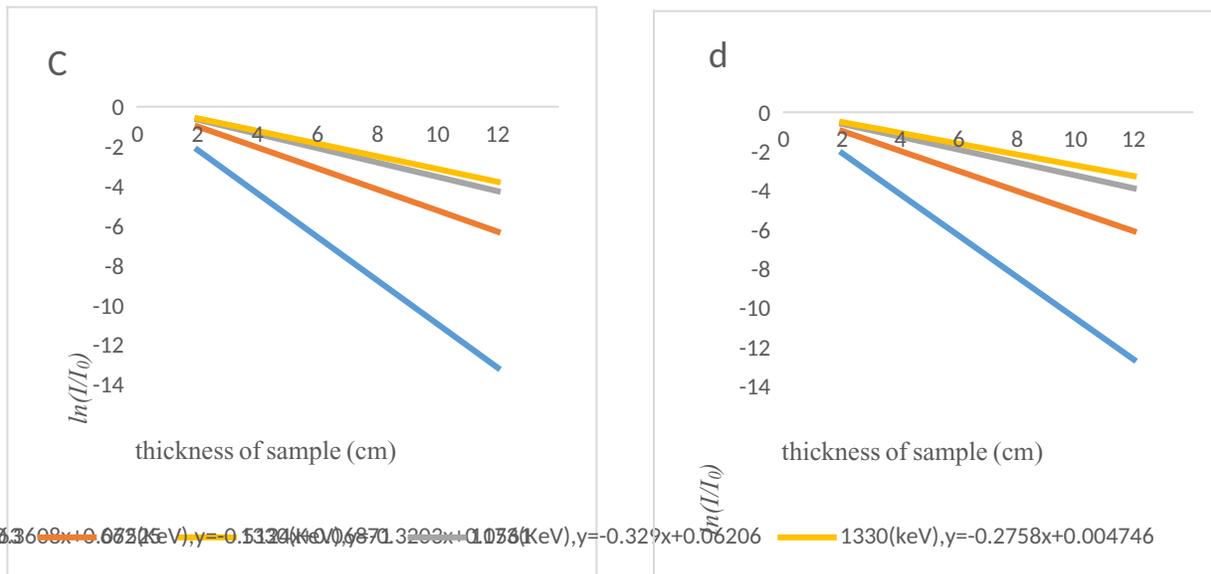

**Fig. 3**. The change of the intensity of gamma photons vs. thickness of the sample for (a) $B_1$, (b) $B_2$, (c) $B_3$, and (d) $B_4$.

Similarly, the linear (and also mass) attenuation coefficient of other materials can also be obtained from the equations given in the Tables 5 (a-d). Moreover, the calculated values of the MFP for different glasses sample vs. energy (energies between 200- 1500 keV) are shown graphically in the Fig. 2 (a-d). The best-fitted equation for various samples has been displayed in these figures and Table 6. As can be seen, the variation of mass attenuation coefficient against gamma-ray energy is a first-order exponential function. These equations can reproduce the MFP values of the chosen glass sample (in this study) at any requested energy in the range of 200–1500 keV. For example, the calculated equation, by some useful energy values (in therapeutic range) such as 427.9 keV ($^{125}$Sb), 795.8 keV($^{134}$Cs), 834.8 keV($^{54}$Mn), and 1112.1 keV($^{152}$Eu) have been estimated, and the values of MFP from the fitted equations are obtained. To validate these equations, the RPE between the results of simulation using the MCNPX and the XCOM values was calculated. It was found that the equations obtained from the fitted equations are in good agreement with the results of the simulation and XCOM data on the all energies. Therefore, it can conclude that these equations are suitable for faster and without complexity estimation of photon attenuation coefficient of $B_2O_3$-$Bi_2O_3$-ZnO-$Li_2O$ based on the glass samples in the energies between the 200–1500 keV.

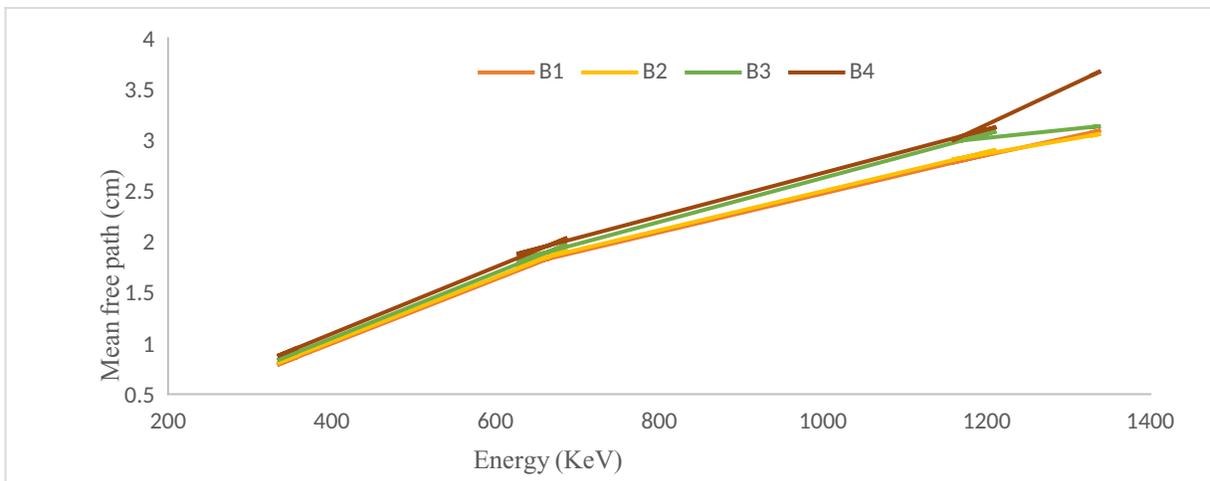

**Fig. 4.** Comparison of the mean free path of different glass systems vs. photon energy.

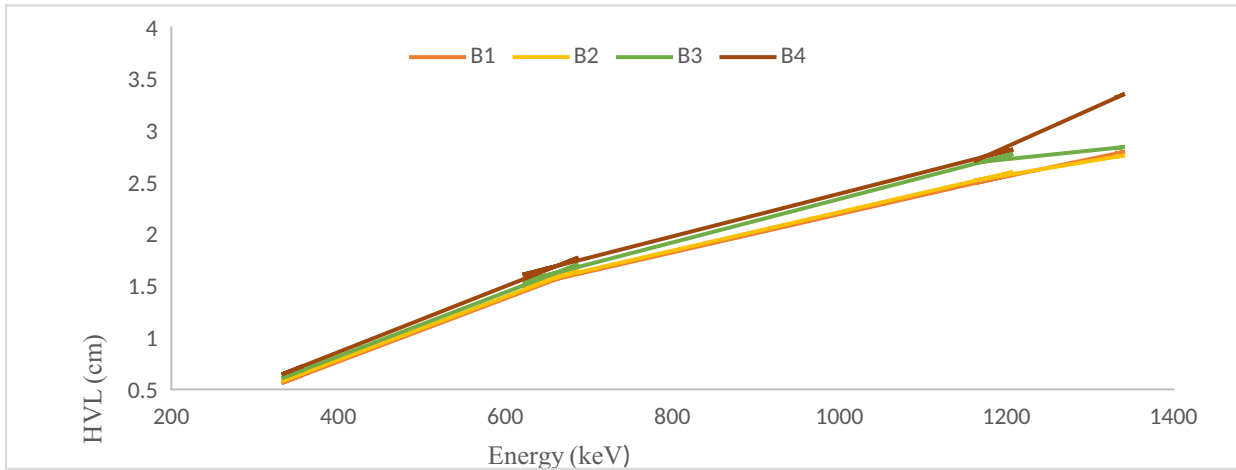

**Fig. 5.** Comparison of the half-value layer of different glass systems vs. photon energy.

After calculation of HVL and MFP parameters, the efficacy of materials in reducing the radiation is directly and briefly described. As it can be seen, more radiation protection performance can be achieved by the materials with low amount of HVL and MFP parameters. Therefore, both of the MFP and HVL values of $B_1$– $B_4$ glasses were determined. The changes of MFP and HVL parameters vs. energy for different glass samples are shown in Figs. 4 and 5, respectively. As shown, the values of MFP and HVL increase vs. photon energy. Accordingly, photons with higher energy have a higher penetration power in the samples under study than lower energy photons, which are physically consistent with what is seen in the attenuation coefficient diagram. On the other hand, the MFP and HVL values of samples are in the descending order of $B_4 > B_3 > B_2 > B_1$. From these results, shielding parameters of these samples are inversely related to the density of matter (glass samples). Therefore, among the samples studied in this study, case $B_4$ provides the best shielding effect against gamma-ray.

**Table 6.** The proposed equations for estimation of mass attenuation and comparison to results of performed simulation by MCNP computer code.

| Glass sample | Best Fitted Equation | Energy(keV) | Fitted Equation | MCNP | Percentage |
|---|---|---|---|---|---|
| $B_1$ | y=2.469exp(-0.0079x)+0.06654 | 427.9($^{125}$Sb) | 0.1505 | 0.1491 | 0.009 |
| $B_2$ | y=2.5969exp(-0.0081x)+0.06667 | 795.8($^{134}$Cs) | 0.0707 | 0.0777 | 0.09 |
| $B_3$ | y=2.227exp(-0.0074x)+0.06265 | 834.8($^{54}$Mn) | 0.06713 | 0.07012 | 0.04 |
| $B_4$ | y=2.457exp(-0.0079x)+0.06226 | 1112.1($^{152}$Eu) | 0.06226 | 0.06157 | 0.017 |

## 4. Conclusions

As mentioned in the present paper, adding lead to any substance causes toxicity in the nature of that substance, so excessive use of lead is not appropriate. On the other hand, very cheap and available concrete materials are widely used to protect against gamma-rays. However, such materials have major problems: (1) the concrete is opaque to visible light, (2) they are difficult to move due to their high weight. Today, glass materials are receiving more attention as new and better shielding performance. The composition studied in this study, due to its appropriate transparency against visible light and the ability to pass light, appropriate weight in comparison to the concrete and lead and acceptable protective function against gamma-rays is a good choice for photon energy range used in radiotherapy.

In this study, the MCNPX computer code was used to investigate the important parameters of radiation shielding of 4 compositions of $B_2O_3$-$Bi_2O_3$-ZnO-$Li_2O$ mixture with different mole percentage of any component. The effect of different mole percentage of $Li_2O$ on shielding performance of the glass sample was investigated. To this end, the main parameters in radiation protection including attenuation coefficient, HVL, and MFP of four diffrent glass samples were calculated. The results of the simulations were validated against the XCOM data. It is found that there is a good agreement between results from the MCNPX computer code and XCOM database.

For some useful energy values, the credibility of the models create in the MCNPX computer program has been studied. The relative percentages difference between the $\mu_m$ values of MCNPX computer code and XCOM database for $B_2O_3$-$Bi_2O_3$-ZnO-$Li_2O$ system has been determined. We have a RPE of less than 4% for all energies except energy 200. In addition, HVL and MFP results show that the case $B_4$ has the highest shielding effect among the studied samples.

**Acknowledgment**

The authors are grateful to the research office of the Sharif University of Technology for the support of the present work.